%% file: main.tex
\documentclass[twocolumn,english,10pt,aps,prd,a4paper,preprintnumbers,floatfix,nofootinbib,showpacs,superscriptaddress, notitlepage]{revtex4-1}

\usepackage{graphicx}
\usepackage{amsmath}
\usepackage{amssymb}
\usepackage{amsfonts}
\usepackage{physics}
\usepackage{dcolumn}
\usepackage{bm}
\usepackage[dvipsnames]{xcolor}
\usepackage[linktoc=none]{hyperref}

\usepackage{braket}
\usepackage{tabstackengine}
\stackMath
\usepackage[mathscr]{euscript}

\begin{document}

\title{Fermionic sub-GeV Dark Matter from evaporating Primordial Black Holes at DarkSide-50}

\input{authors_DS50}
\input{inst}

\begin{abstract}
We present a search for boosted dark matter from Primordial Black Holes (PBH) evaporation using the DarkSide-50 ionization-signal-only dataset corresponding to the experiment's ($12202\pm180$) 
${\rm kg\: d}$ exposure. We focus on evaporation of PBHs with masses in the range [$10^{14},\,10^{16}$] g producing Dirac fermionic dark matter particles with sub-GeV kinetic energy. 
These relativistic particles, with energies up to hundreds of MeV, can generate detectable signals for masses below $\mathcal{O}(100)$ MeV. The absence of a signal enables setting complementary limits to those derived from cosmological observations and direct detection searches for cosmic ray-boosted dark matter.
\end{abstract}


\maketitle
Astrophysical and cosmological observations over a wide range of length scales suggest the presence of Dark Matter (DM) in the Universe\,\cite{Bertone:2018krk, Cirelli:2024ssz}.
Despite being regarded as one of the backbones of the standard cosmological model, we have almost no information regarding its nature and interaction, apart from its gravitational effects. Among the detection techniques employed to discern DM properties, one finds direct detection experiments looking for DM-nucleon and DM-electron interactions\,\cite{Schumann:2019eaa}. The drawback of looking at the se interactions is that the recoil energies induced by sub-GeV DM,\footnote{Natural units $c=\hbar=1$ are used throughout this paper.} predicted by several mechanisms\,\cite{Essig:2013lka, Battaglieri:2017aum}, are below the detection threshold of most direct detection experiments. Scenarios predicting a fraction of DM endowed with high kinetic energies overcome this experimental limitation\,\cite{Agashe:2014yua, Giudice:2017zke, Fornal:2020npv, Cappiello:2018hsu, Bringmann:2018cvk, Ema:2018bih, Cappiello:2019qsw, Ema:2020ulo, PROSPECT:2021awi}. 
In this paper, we focus on detecting boosted sub-GeV DM from evaporating Primordial Black Holes (PBHs) using data from the DarkSide-50 experiment. 

The DarkSide-50 experiment is a dual-phase liquid argon (LAr) time projection chamber (TPC). It operated underground in Italy at the INFN Gran Sasso Laboratories (LNGS). 
Energy released by particle interactions in the 46.4$\pm$0.7 kg active LAr volume results in two measurable quantities: scintillation light (S1) and ionization ion-electron pairs. The ionization electrons are drifted by a 200 V/cm electric field towards the upper part of the TPC, where they are extracted into a gas pocket by a stronger 2.8 kV/cm electric field, inducing a secondary electroluminescence light signal (S2)~\cite{Charpak:1975dq, Policarpo:1981vn, DarkSide:2018stg, DarkSide:2020oas, Buzulutskov:2020xhd}. 
The 128 nm ultraviolet S1 and S2 photons are converted into 420 nm photons by a tetraphenyl butadiene wavelength shifter coating the internal surface of the TPC. Below the cathode and above the anode, the 420 nm photons are finally detected by two arrays of 19 3-in photomultiplier tubes (PMTs). 
The TPC is contained in a stainless steel cryostat. The cryostat lies inside a 30 t boron-loaded liquid scintillator equipped with 110 8-in PMTs and is used as an active neutron veto. 
The neutron veto is finally enclosed in a 1 kt ultra-pure water Cherenkov tank equipped with 80 PMTs that provides a passive shield against external backgrounds and that is used as a cosmic muon veto. 
Additional information on the DarkSide-50 detector is given in Refs.~\cite{DarkSide:2014llq,DarkSide:2015cqb,DarkSide:2018bpj,DarkSide:2018kuk,DarkSide:2018ppu}.

The DarkSide-50 experiment has searched for DM with masses ranging from hundreds of MeV to several GeV with the ionisation-only and Migdal analyses~\cite{DarkSide-50:2022qzh,DarkSide:2022dhx}. Dark matter particles with lower masses could still be detectable by DarkSide-50 if they had sufficient kinetic energy. Evaporating PBH could provide such a mechanism, as shown in Ref.~\cite{Calabrese:2021src, Calabrese:2022rfa}. PBHs, hypothetical black holes generated as a byproduct of some inflationary (or more complex) scenarios\,\cite{Zeldovich:1967lct, Harrison:1969fb}, offer a broader mass range than ordinary black holes because of their non-stellar origin. Hawking's studies on the quantum effects of PBHs revealed that they evaporate, emitting elementary particles with masses smaller than the PBHs' temperature\,\cite{Hawking:1974rv, Hawking:1975vcx}. This unique property places several constraints on PBHs abundance \cite{Boudaud:2018hqb, Laha:2019ssq, Ballesteros:2019exr, Dasgupta:2019cae, Coogan:2020tuf, Calabrese:2021zfq, DeRomeri:2021xgy, Baker:2021btk, Carr:2020gox, Saha:2021pqf, Saha:2024ies}. 

According to Hawking's calculations, spinless neutral PBHs with masses exceeding $5\times 10^{14}$ g survive up to the present time.
Many phenomenological studies focused on the production of DM from PBH evaporation in the early Universe\,\cite{Morrison:2018xla, Baldes:2020nuv, Bernal:2020kse, Bernal:2020ili, Bernal:2020bjf, Auffinger:2020afu, Gondolo:2020uqv, Masina:2021zpu, Cheek:2021odj, Cheek:2021cfe}. In contrast, we consider the production of DM up to the present time. Moreover, the generated DM constitutes a small contribution to the current DM abundance\,\cite{Calabrese:2021src, Calabrese:2022rfa, CDEX:2022dda, CDEX:2024xqm}. However, its presence yields interesting implications for direct detection facilities. The analysis implemented in this paper leads to constraints in the DM parameter space that are complementary with existing ones~\cite{Bringmann:2018cvk,Cappiello:2019qsw, PandaX-II:2021kai,CRESST:2017ues,CRESST:2019jnq}.


In our search strategy, we characterize the DM flux from PBHs with a monochromatic mass distribution. We consider non-rotating and neutral PBHs, as spinning PBHs exhibit an enhanced evaporation rate. Within this context, the Hawking temperature is\,\cite{Hawking:1974rv, Hawking:1975vcx, Zeldovich:1976vw, Carr:1976zz, Page:1976df, Page:1977um, MacGibbon:1990zk}:
\begin{equation}
\mathcal{T}_{\rm PBH} = 10.6 \left(\frac{10^{15}\,{\rm g}}{{\rm M}_{\rm PBH}}\right)\,{\rm MeV}\,,
\end{equation}
where ${\rm M}_{\rm PBH}$ is the PBH mass in in grams. The differential spectrum of emitted fermion DM species $\chi$ from a PBH with Hawking temperature $\mathcal{T}_{\rm PBH}$ reads
\begin{equation}
\frac{\dd^2 N_\chi}{\dd T_\chi\dd t} = \frac{g_\chi}{2\pi} \frac{\Gamma^\chi(T_\chi, \mathcal{T}_{\rm PBH})}{\exp{(T_\chi+m_\chi)/\mathcal{T}_{\rm PBH})}+ 1}\,,
\end{equation}
where $m_\chi$, $T_\chi$ and $g_\chi$ are the particle's mass, kinetic energy, and degrees of freedom, respectively. This expression accounts for deviation from the black-body radiation by means of the grey-body factor $\Gamma^\chi$, obtained from \texttt{BlackHawk}\,\cite{Arbey:2019mbc, Arbey:2021mbl, Auffinger:2022sqj}, a code that computes the Hawking emission. We assume our DM candidate to be a Dirac fermion ($g_\chi = 4$). 

After estimating the differential spectrum, we quantify the flux of light DM reaching the Earth. We divide the flux of DM particles from PBHs into galactic (gal.) and extragalactic (egal.) components
\begin{equation}
    \frac{\dd^2 \phi_\chi}{\dd T_\chi\dd \Omega} = \frac{\dd^2 \phi_\chi^{\rm gal.}}{\dd T_\chi\dd \Omega} + \frac{\dd^2 \phi_\chi^{\rm egal.}}{\dd T_\chi\dd \Omega}\,.
\end{equation}
We describe the galactic component as 
\begin{equation}
    \frac{\dd^2 \phi_\chi^{\rm gal.}}{\dd T_\chi\dd \Omega} = \frac{f_{\rm PBH}}{4\pi {\rm M}_{\rm PBH}} \frac{\dd^2 N_\chi}{\dd T_\chi \dd t} \int_0^{+\infty}\dd s \rho_{\rm NFW}(r)\,,
\end{equation}
where $f_{\rm PBH}\equiv\rho_{\rm PBH}/\rho_{\rm DM}$ is the fraction of the  DM energy density composed of PBHs in the current cosmic landscape. The galactic component is proportional to the integral over the line-of-sight distance $s$ of the galactic DM density. Our analysis adopts the Navarro-Frank-White DM density profile $\rho_{\rm NFW}$\,\cite{Navarro:1996gj}, that is a function of the galactocentric distance $r = (r_\odot^2 +s^2 -2r_\odot s\cos \ell \cos b)^{\frac12}$, where $r_\odot = 8.5$\,kpc is the distance between the sun and the galactic centre, and $(b,\,\ell)$ are the galactic latitude and longitude. 

\begin{figure}
    \centering
    \includegraphics[width = .48\textwidth]{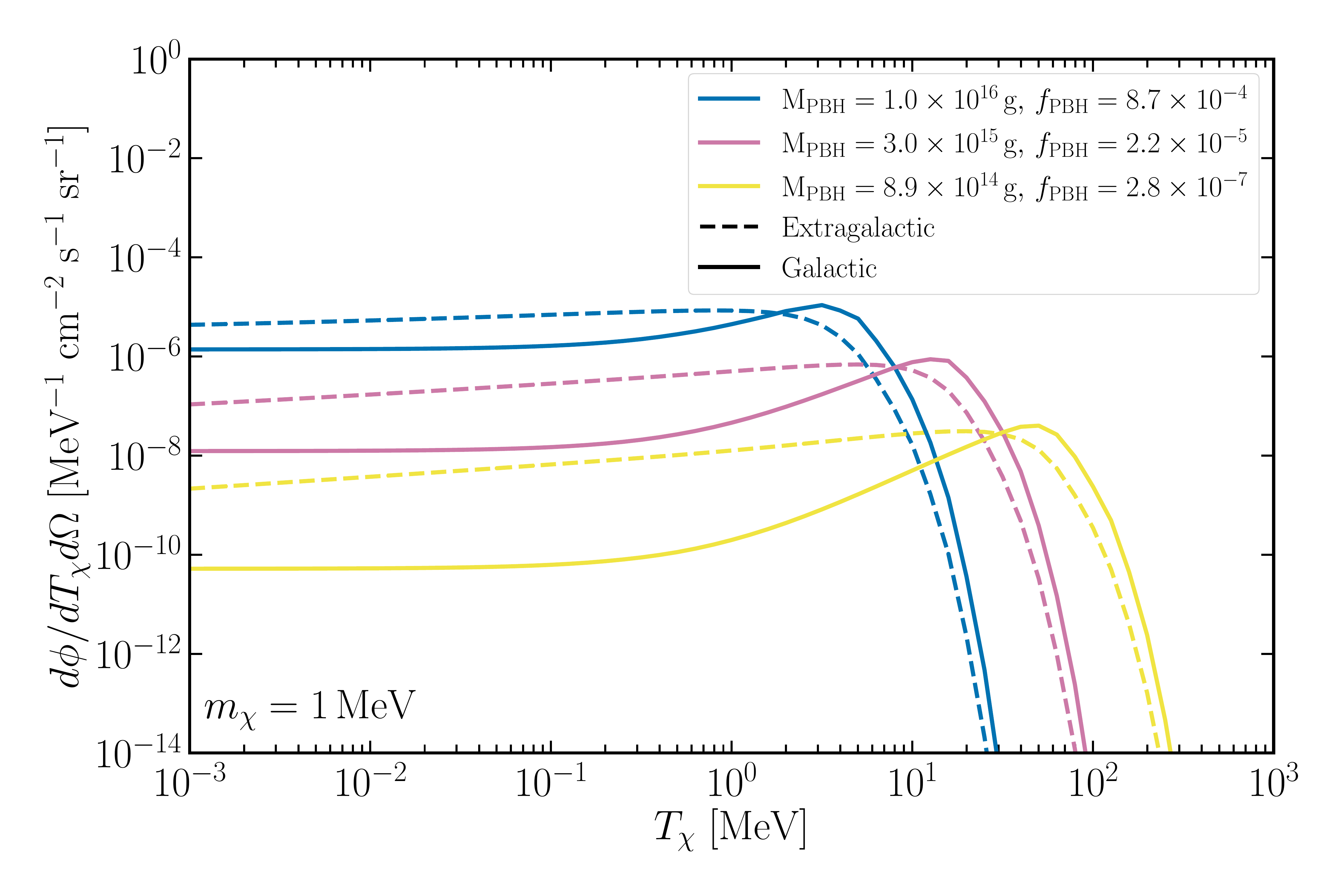}
    \caption{Diffuse DM differential flux from evaporating PBHs as a function of its kinetic energy. The solid lines show the galactic components, while the dashed lines the extragalactic ones. With different colors, we identify different PBH masses and abundances.
    }
    \label{fig:fluxes}
\end{figure}

We consider the following expression for the extragalactic differential flux 
\begin{equation}
 \frac{\dd^2 \phi_\chi^{\rm egal.}}{\dd T_{\chi}\dd \Omega} = \frac{f_{\rm PBH}\rho_{\rm DM}}{4\pi {\rm M}_{\rm PBH}}\int_{t_{\rm min}}^{t_{\rm max}} \dd t [1+z(t)] \frac{\dd^2 N_\chi}{\dd T_{\chi} \dd t}\,,
\end{equation}
where $z(t)$ is the redshift, and we perform the integral from the time of matter-radiation equality ($t_{\rm min}$) to the age of the Universe ($t_{\rm max}$). As $z(t)$ increases, the contribution of boosted dark matter becomes progressively negligible due to the redshift. In contrast with the galactic component, which is enhanced towards the galactic center, the extragalactic one is isotropic. In Fig.~\ref{fig:fluxes}, we show the galactic and extragalactic diffuse flux for a reference DM mass of $m_\chi = 1$\,MeV for different PBH masses assuming the largest allowed $f_{\rm PBH}$ by constraints of Refs.~\cite{Carr:2020gox, Iguaz:2021irx}. 

Direct detection facilities can detect DM via their interaction with nucleons and electrons.
However, these processes are not only fundamental for detection but also for the attenuation of the DM flux at the DarkSide-50 due to propagation through the Earth and its atmosphere\,\cite{Starkman:1990nj, Mack:2007xj, Kavanagh:2016pyr, Emken:2018run}.

In this analysis, we adopt the analytical approach described in Ref.~\cite{Bringmann:2018cvk}, where the authors consider a ballistic DM propagation. However, as shown in Ref.\,\cite{Cappiello:2019qsw}, this approximation would not hold for light DM collisions with heavy nuclei. 
Instead, they consider a diffusive propagation that may lead to a larger energy loss, a possible reflection of DM, and a lower flux. To better understand the impact of the ballistic approximation, we also estimate $\chi$ flux under the most conservative assumption that any interaction depletes the particles completely. 
Even under this assumption, the expected event rate in DarkSide-50 is still detectable, as the difference in event rates is around 12\%.. Because of this, we do not expect that a more detailed model of the propagation of a light DM candidate would substantially alter our scenario, we simply use the ballistic approach. 

For PBHs with masses greater than $5\times10^{14}$ g, the Hawking temperature -- and therefore the dark matter kinetic energy -- remains below $100$ MeV. Consequently, the impact of the nuclear form factor in the propagation through the Earth, as well as the effects of quasi-elastic and inelastic scattering, can be safely neglected~\cite{Alvey:2022pad}.
Within this approximation, the interaction length for propagation through Earth ($\ell^\oplus_\mathrm{int}$) and the atmosphere ($\ell^{\rm atm}_\mathrm{int}$) are
\begin{equation}
    \ell^i_\mathrm{int}= \left[\sum_\mathcal{N} n^i_\mathcal{N} \sigma_{\chi\mathcal{N}} \frac{2m_\mathcal{N} m_\chi}{(m_\mathcal{N} +m_\chi)^2}\right]^{-1}\,,
\end{equation}
where the sum is taken over the most abundant elements in the Earth and in the atmosphere, $\mathcal{N}~=~\{\text{Fe}, \text{Al}, \text{N}, \text{O}, \text{Si}, \text{Ni}, \text{S}, \text{Mg}, \text{Ca}\}$, $n_\mathcal{N}$ is their abundance and $m_{\mathcal{N}}$ their mass.

Assuming $\chi$ couples equally to protons and neutrons -- a common benchmark in searches for WIMPs -- the DM-nucleon cross-section is
\begin{equation}
    \sigma_{\chi\mathcal{N}}= \sigma_\chi^{\mathrm{SI}} \mathrm{A}^2 \left(\frac{m_\mathcal{N} (m_\chi+m_\mathrm{p})}{m_\mathrm{p}(m_\chi +m_\mathcal{N})}\right)^2 \,.
\end{equation}
where $A$ is the $\mathcal{N}$ mass number, $m_p$ is the proton mass, and $\sigma_\chi^{\mathrm{SI}}$ denotes the spin-independent cross-section, the focus of our analysis. We highlight that spin-independent scattering is subject to a coherent enhancement. We indicate with $T_0$ the initial kinetic energy of a particle before entering the atmosphere, and with $T_d$ the kinetic energy of a dark matter particle that traveled a total distance $d = d_{\rm atm} + d_\oplus$, where $d_{\rm atm}$ is the thickness of the atmosphere traversed by the DM particle, and $d_\oplus$ the depth of the DarkSide-50 experiment. The distances $d_{\rm atm}$ and $d_\oplus$ depend on the time-dependent longitude of the experiment.
We account for the DarkSide-50 detector located at a depth of about $1.4~{\rm km}$~\cite{capuano1998density}. The flux of DM particles in DarkSide-50 is 
\begin{equation}
\label{eq:att}
    \frac{\dd^2 \phi_\chi^{d}}{\dd T_{d} \dd \Omega}\approx\frac{4m_\chi^2e^\tau}{\left(2m_\chi + T_d - T_d e^\tau\right)^2}\left(\frac{\dd^2 \phi_\chi}{\dd T_\chi \dd \Omega}\bigg|_{T_0}\right)\,,
\end{equation}
where $\tau = d_{\rm atm}/\ell^{\rm atm}_{\rm int} + d_\oplus/\ell^\oplus_{\rm int}$, and the incoming DM flux is evaluated at
\begin{equation}
    T_0\left(T_d\right)=\frac{2m_\chi T_d\,e^\tau}{2m_\chi + T_d - T_d e^\tau}\,.
\end{equation}
We consider the average of Eq.~\ref{eq:att} over a day.

After estimating the flux of DM at the detector, we can calculate the differential event rate. It quantifies the number of expected events at the detector due to the interaction of DM with the target material. In the case of DM-nucleon interaction, the differential event rate reads as 
\begin{equation}
\frac{\dd R}{\dd E_r} = \sigma_{\chi Ar} \mathcal{N}_{Ar} \int \dd T_d \dd \Omega \frac{\dd^2 \phi^d_\chi}{\dd T_d \dd \Omega} \frac{\Theta(E_r^{\rm max}- E_r)}{E_r}\,,
\end{equation}
where $\mathcal{N}_{Ar}$ is the total number of targets, $\Theta$ is the Heaviside function, $E_r$ is the recoil energy, and 
\begin{equation}
    E_r^{\rm max} = \frac{T_d^2 + 2 m_\chi T_d}{T_d + (m_\chi + m_{Ar})^2/(2 m_{Ar})}
\end{equation}
is the maximum recoil energy allowed,with $m_\mathrm{Ar}$ being the argon mass. The spin-independent cross-section $\sigma_{\chi Ar}$, including also the argon nuclear form-factor~\cite{Helm:1956zz, Lewin:1995rx}, is defined as 
\begin{equation}
    \sigma_{\chi Ar} = \sigma_\chi^{SI} A^2 \left(\frac{m_N(m_\chi + m_p)}{m_p (m_\chi + m_N)}\right)^2F(q)^2\,. 
\end{equation}
where $q$ is the three-momentum transfer and $F(q)$ is the Helm nuclear form factor.

\begin{figure}
    \centering
    \includegraphics[width = .48\textwidth]{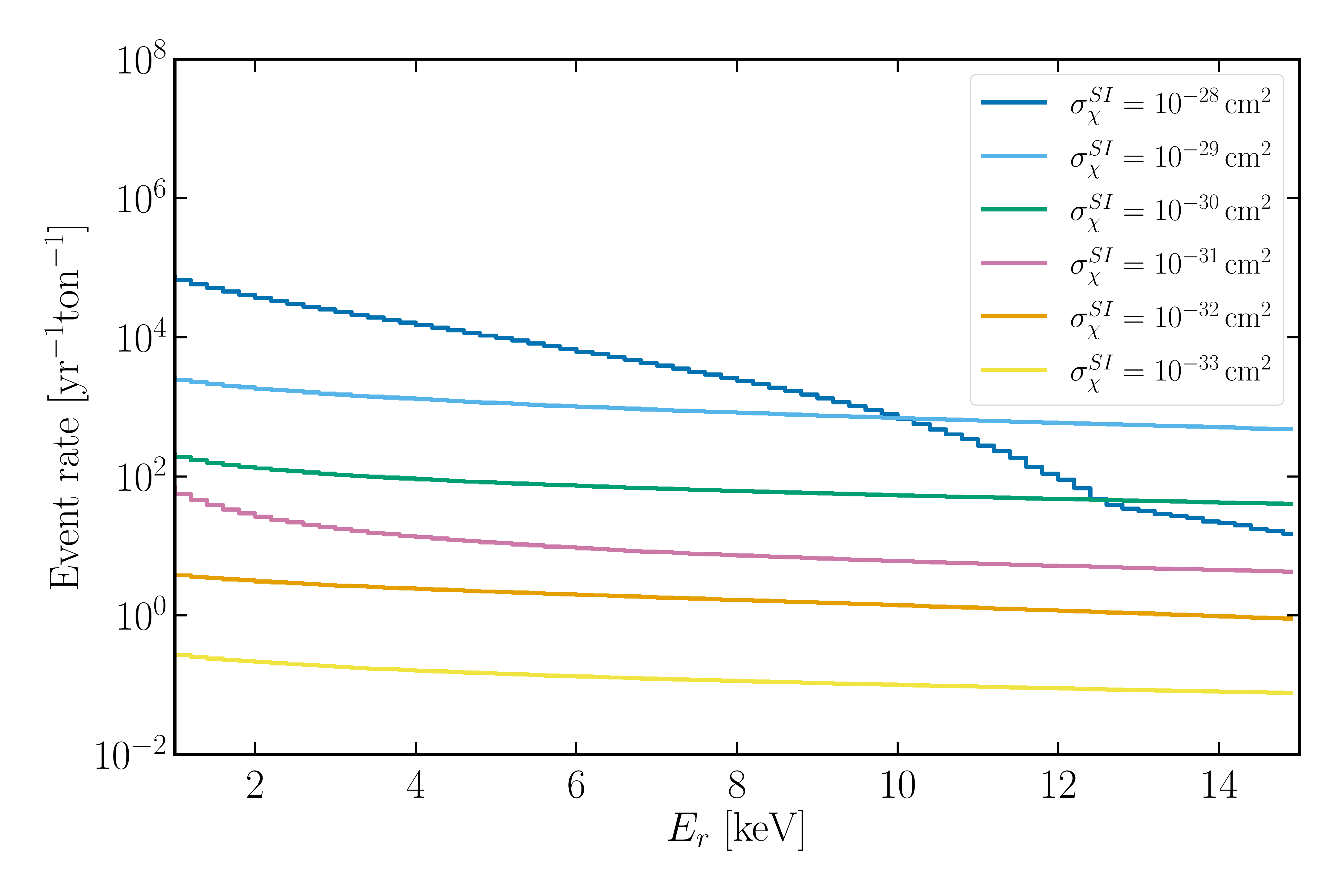}
    \caption{
    Expected rate assuming M$_{\rm PBH} = 1.0\times10^{15}\,$g and $f_{\rm PBH} = 4.1\times10^{-7}$ for different spin-independent cross-sections.}
    \label{fig:RATE}
\end{figure}

In Fig.~\ref{fig:RATE}, we show the expected event rate vs recoil energy assuming M$_{\rm PBH}=1.0\times10^{15}\,$g and the highest $f_{\rm PBH} = 4.1\times10^{-7}$ compatible with existing limits~\cite{Carr:2020gox, Iguaz:2021irx}. Initially, the rate increases for higher cross-sections, until attenuation effects become significant and shift the spectrum toward lower recoil energies. As the DM lose energy traveling through the Earth’s overburden, they arrive at the detector with reduced kinetic energy, which leads to a suppression of high-energy recoils but an enhancement of the rate at lower recoil energies. For even larger cross-sections, this enhancement moves to recoil energies below the detector’s threshold, ultimately causing the observed rate to decrease overall.

For the analysis described in this work, we consider only the S2 signal, regardless of the presence or absence of an accompanying S1 signal. The advantage of considering the so-called ``S2-only" dataset is that, without requiring S1, the low-energy threshold to nuclear recoils drops significantly. 
However, without S1, there is no possibility of discriminating between electronic and nuclear recoils, therefore requiring a rigorous treatment of the electron recoil background and its related systematic uncertainties. 
A rigorous definition of the background model can be found in Ref.~\cite{DarkSide-50:2022qzh}. The main background contributions are given by ``external" $\gamma$ background events induced by the radioactivity of the PMTs and the cryostat, and by ``internal" background events induced by the decays of $^{85}$Kr and $^{39}$Ar isotopes within the active LAr volume. 
Other contributions, including solar and atmospheric neutrinos and cosmogenic or radiogenic neutrons, are negligible. The $\beta$ decay spectra associated to the $^{85}$Kr and $^{39}$Ar background are computed including the atomic exchange and screening effects and are validated down to $200\:{\rm eV}$~\cite{PhysRevA.90.012501, Haselschwardt:2020iey}: below this threshold we consider an additional symmetric systematic uncertainty on the spectral shape ranging linearly from 25\% at 0 eV to 0\% at $200$~eV~\cite{DarkSide-50:2022qzh}. 
Moreover, the uncertainty on the Q-value~\cite{Wang:2021xhn} is also taken into account as a systematic $1\%$ uncertainty on the shape of the spectrum of $^{39}$Ar and $0.4\%$ on the one of $^{85}$Kr. The other systematic effect acting on the shape of all the background spectra being considered is the uncertainty on the ionization response, modeled via Monte Carlo simulations and constrained by calibration measurements~\cite{DarkSide:2021bnz}. 
The external background spectra are computed by means of Monte Carlo simulations of the radioactivity of each component of the detector. 
The simulations use the initial activities and their associated uncertainties from material radiopurity measurements and account for their decay over the data taking period.

The dataset considered in this paper corresponds to the total $(12202 \pm 180)$ kg d exposure of the experiment, and we used the same data selection described in Ref.~\cite{DarkSide-50:2022qzh}. The analysis of the sensitivity of DarkSide-50 to a PBH induced DM signal is carried out with the Bayesian analysis tool developed and presented in Ref.~\cite{DarkSide-50:2023fcw}. This method integrates the detector response model into the likelihood function, explicitly preserving the dependence on the quantity of interest, $\sigma^{\rm SI}_{\chi}$. In particular, as done in Ref.~\cite{DarkSide-50:2023fcw}, we assume a binned Poisson likelihood defined as
\begin{equation}
    p(\{x_i\} | \boldsymbol{\theta}) = \prod_i
    \frac{\lambda_i(\boldsymbol{\theta})^{x_i}}{x_i !} e^{-\lambda_i(\boldsymbol{\theta})},
    \label{eq:likelihood}
\end{equation}
where $\lambda_i$ is the intensity of the Poisson process in the $i$-th bin of the data spectrum, $x_i$ is the corresponding observed number of events, and $\boldsymbol{\theta}$ indicates all the parameters of the fit related to the signal model and the detector response and background model - i.e. the calibrations, the background rates, and the signal parameters. 
The data spectrum is expressed as a function of the number of detected primary ionization electrons, $N_{e^-}$. This quantity, that is not necessarily an integer, is proportional to S2 -- and consequently to the nuclear recoil energy -- through the parameter $g_2 = (23 \pm 1)\:\text{photoelectrons}/e^-$. This represents the mean number of photoelectrons produced per ionization electron extracted in the gas pocket~\cite{DarkSide:2021bnz}. The lower $N_{e^-}$ threshold for this analysis is set at $N_{e^-}=4$, where the contribution of correlated background events due to spurious single-electrons is negligible. The upper threshold is defined by the maximum value at which detector response calibrations have been validated, specifically $N_{e^-} = 170$~\cite{DarkSide-50:2022qzh, DarkSide:2022dhx, DarkSide-50:2023fcw}.
For this analysis, $\lambda_i(\boldsymbol{\theta})$ is given by
\begin{align}
    \lambda_i =\frac{\mathcal{E}}{\mathcal{E_{\rm DS}}} \:\Big[&r_{B, {\rm Ar}} \; S_i^{\rm Ar}(\boldsymbol{\theta}_{nuis}) + \nonumber \\
    &r_{B, {\rm Kr}} \; 
    S_i^{\rm Kr}(\boldsymbol{\theta}_{nuis}) +\nonumber\\
    &r_{B, {\rm PMT}} \; S_i^{\rm PMT}(\boldsymbol{\theta}_{nuis}) +\nonumber \\ 
    &r_{B, {\rm cryo}} \; S_i^{\rm cryo}(\boldsymbol{\theta}_{nuis})  +\nonumber \\
    &\; S_i^{\rm \chi}(\sigma^{\rm SI}_{\chi}, \boldsymbol{\theta}_{nuis}, f_{\rm PBH} ) 
   ] \,,
    \label{eq:lambda}
\end{align}
where $S_i^{src}$, with $src \in \{\mathrm{Ar}, \mathrm{Kr}, \mathrm{PMT}, \mathrm{cryo}, \mathrm{\chi}\}$, are the expected background and signal spectra for the nominal total exposure $\mathcal{E_{\rm DS}}$ as a function of the detector response nuisance parameters $\boldsymbol{\theta}_{nuis}$ and the cross-section $\sigma^{\rm SI}_{\chi}$. 
The variables $r_{B, src}$ are proportional to the rate of the internal and external background components. Here we use the same normalization of Ref.~\cite{DarkSide-50:2023fcw}, where the spectra are normalized in such a way that $r_{B, src} = 1$ corresponds to the nominal DarkSide-50 exposure $\mathcal{E_{\rm DS}}$. 
The quantity $f_{\rm PBH}$ is not treated as a parameter of the fit but it is fixed, depending on $\mathcal{M}_\mathrm{PBH}$, to the highest value allowed by existing constraints derived from Extra-Galactic gamma-ray data~\cite{Carr:2020gox} and from isotropic X-ray observations~\cite{Iguaz:2021irx}.
For smaller values of $f_{\rm PBH}$, the corresponding upper bound on the cross section derived in this work can be obtained as $\sigma_\chi^{\rm SI,\:new } = \sigma_\chi^{\rm SI,\:old } f_{\rm PBH}^{\rm old} / f_{\rm PBH}^{\rm new}$. This approximation remains valid until the cross section becomes large enough for Earth shielding effect to significantly limit the detector’s sensitivity. In that regime, the fit must be repeated using the correct signal spectrum.
A rigorous description of the detector response model is described in Ref.~\cite{DarkSide-50:2022qzh}, while its implementation in this analysis framework is given in Ref.~\cite{DarkSide-50:2023fcw}. 
The prior probability density function associated to the $\boldsymbol{\theta}_{nuis}$ and the $r_{B, src}$ parameters corresponds to the constraints coming from the calibration, while a flat prior in the  $[0, 10^{-28}]\:{\rm cm}^2$ range is used for $\sigma^{\rm SI}_{\chi}$. The quantity $\mathcal{E}=(12202 \pm 180)$ kg d represents the total exposure, with a prior uncertainty of 1.5\%~\cite{DarkSide-50:2022qzh}.

\begin{figure}
    \centering
    \includegraphics[width = 0.48\textwidth]{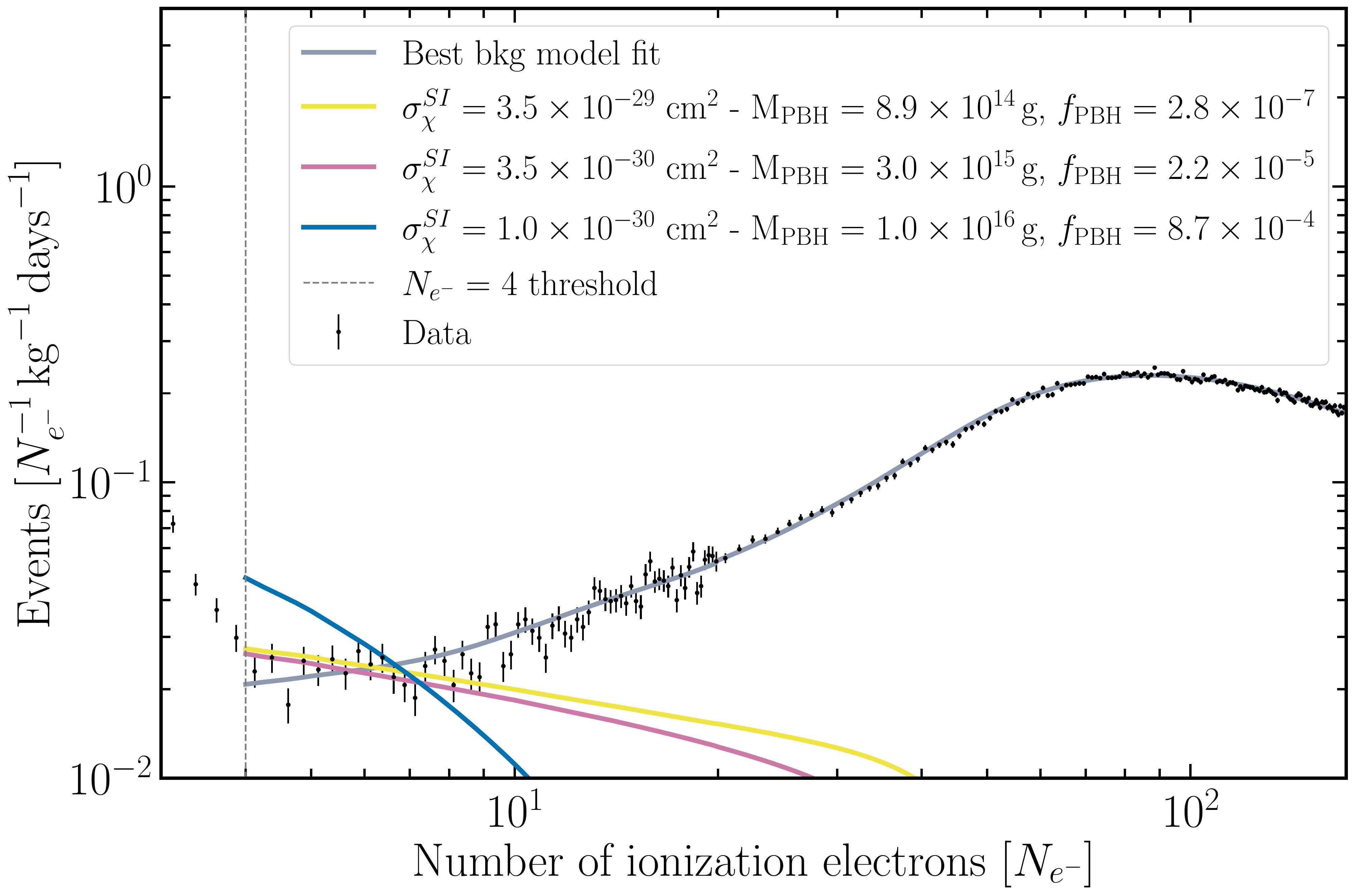}
    \caption{The data spectrum (black points) and the best background-only fit result (gray line). As an example, signal spectra for three different values of $\sigma^{\rm SI}_{\chi}$ and M$_{\rm PBH}$ at $m_{\chi} = 1\:{\rm MeV}$ are shown by the yellow, purple, and blue lines. As shown in Fig.~\ref{fig:fluxes}, lighter PBHs result in harder DM spectra.}
    \label{fig:spectra}
\end{figure}

\begin{figure}
    \centering
    \includegraphics[width = .48\textwidth]{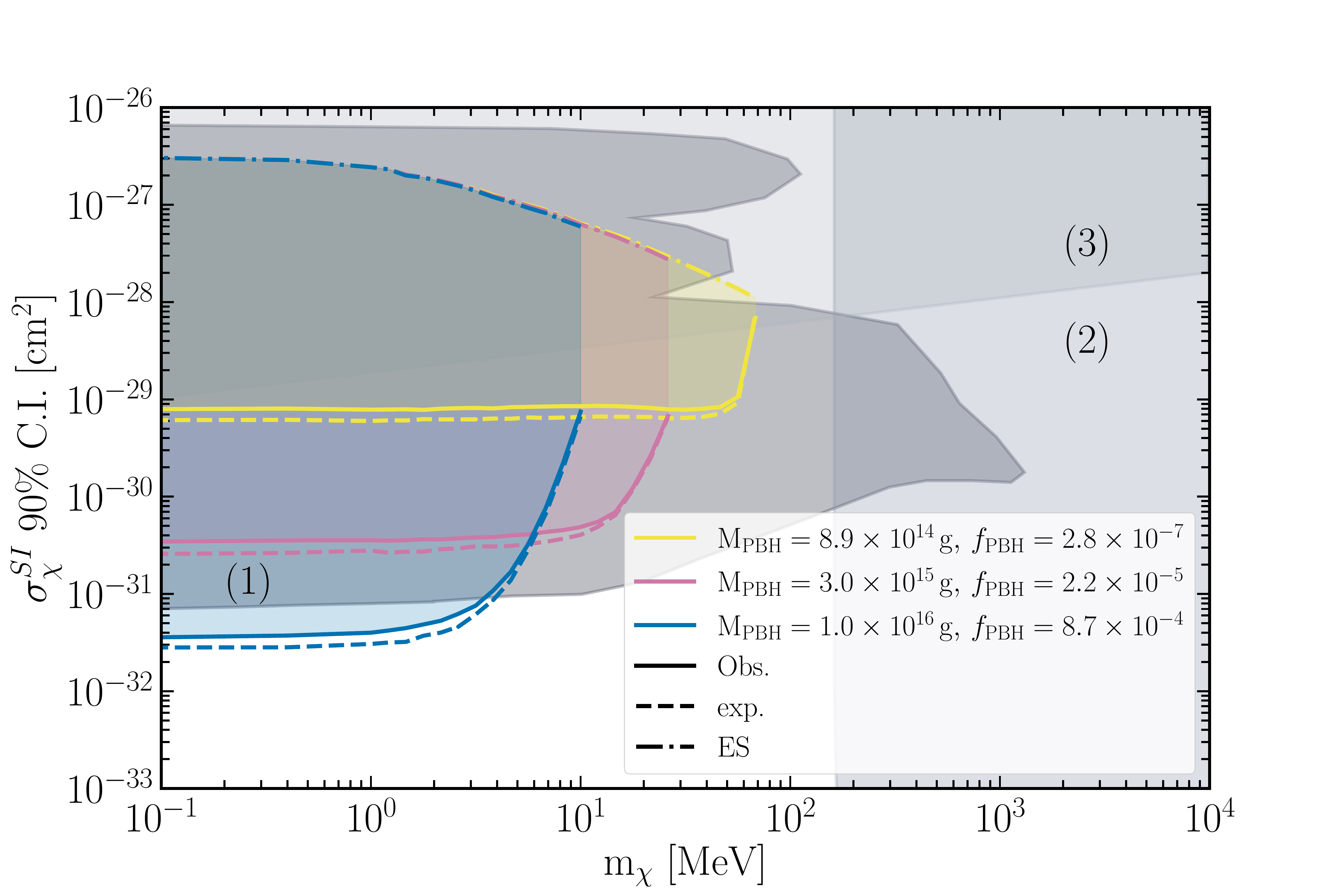}
    \caption{
    The 90\% C.I. observed (solid lines) and expected (dashed lines) limit on the DM cross-section $\sigma^{\rm SI}{\chi}$ as a function of the DM mass, assuming a Dirac fermion DM candidate, and for three different values of the PBH mass for the final DS-50 exposure of 12202 $\pm$ 180 kg d. The parameter $f_\mathrm{PBH}$ is fixed by the constraints of Refs.~\cite{Carr:2020gox,Iguaz:2021irx} (see the text for further details). For comparison, we report DM constraints from (1) cosmic-ray boosted DM particles~\cite{Bringmann:2018cvk,Cappiello:2019qsw, PandaX-II:2021kai}, (2) from the CRESST experiment (in the standard halo model scenario)~\cite{CRESST:2017ues,CRESST:2019jnq}, (3) from cosmology~\cite{Gluscevic:2017ywp,Xu:2018efh,Slatyer:2018aqg,Nadler:2019zrb}, and (4) from CDEX-10~\cite{CDEX:2022dda}, whose results were obtained in a similar scenario.}
    \label{fig:limits}
\end{figure}

The $\sigma^{\rm SI}_{\chi}$ cross-section upper bound is computed as the so-called 90\% Credible Interval (C.I.), defined as the 90\% quantile of the marginalized posterior probability density function for $\sigma^{\rm SI}_{\chi}$. 
As done in Ref.~\cite{DarkSide-50:2023fcw}, the spectrum below $N_{e^-} < 20$ is binned into $0.25\,N_{e^-}$ bins and the spectrum with $N_{e^-} \geq 20$ uses $1\,N_{e^-}$ bins.
This choice is demonstrated to have a 5–10\% beneficial impact on the sensitivity with rispect to a $1\,N_{e^-}$ binning on the whole spectrum when the signal rapidly falls with $N_{e^-}$~\cite{DarkSide-50:2023fcw}. The Earth shielding (ES) upper limit is defined as the value of the cross-section at which the number of signal events above the $N_{e^-}=4$ threshold goes to zero\footnote{Different definitions for the Earth shielding curve were tested, but given the rapidity with which, due to the Earth shielding effect, the spectrum above threshold falls to zero as the cross section increases, they are not appreciably different from the smallest value of $\sigma^{\rm SI}_{\chi}$ at which there is no event above threshold.}. In Fig.~\ref{fig:spectra} the data spectrum (black points), the best background-only fit result (gray line), and examples of signal spectra (blue, red and orange lines) are reported. These spectra include all the smearing effects due to the detector response and are expressed as a function of $N_{e^-}$.
Figure~\ref{fig:limits} shows the 90\% C.I. expected (dashed lines) and the observed (continuous lines) limits on the $\sigma^{\rm SI}_{\chi}$ cross-section as a function of the DM mass. The expected limits are obtained on the so-called Asimov dataset, namely the dataset generated using the nominal pre-fit best estimation of the $\boldsymbol{\theta}_{nuis}$, $r_{B, src}$, and $\mathcal{E}$ parameters. The observed limits are instead obtained on the real DarkSide-50 dataset. The difference between the two curves is due to the difference between the Asimov dataset and the real dataset, reflecting the variation of the pre- and post-fit expected values for $\boldsymbol{\theta}_{nuis}$, $r_{B, src}$, and $\mathcal{E}$. However, all these variations are compatible with the prior expectations within calibration uncertainties~\cite{DarkSide-50:2022qzh, DarkSide-50:2023fcw}.
\hfill \break

In conclusion, we report on the search for boosted DM from PBH evaporation using the DarkSide-50 S2-only dataset. The analysis was performed in the Bayesian framework presented in Ref.~\cite{DarkSide-50:2023fcw}, using the same dataset and the same detector and background model of the analyses presented in Refs.~\cite{DarkSide-50:2022qzh, DarkSide:2022dhx, DarkSide:2022knj, DarkSide-50:2023fcw}. The constraints in Fig.~\ref{fig:limits} extend up to $M_{\rm PBH} = 1.0\times10^{16}\:{\rm g}$ and represent a complementary result with respect to the cosmological~\cite{Gluscevic:2017ywp,Xu:2018efh,Slatyer:2018aqg,Nadler:2019zrb} and direct detection searches~\cite{CRESST:2017ues,CRESST:2019jnq}.

\textbf{Acknowledgements --}This work has been supported by the Italian grant 2022JCYC9E (PRIN2022) by the Italian Ministero dell’Università e della Ricerca (MUR) within the EU-funded project PNRR M4 ‐ C2 – Inv. 1.1., by the research project TAsP (Theoretical Astroparticle Physics) funded by the Istituto Nazionale di Fisica Nucleare (INFN), and the research grant number 2022E2J4RK ``PANTHEON: Perspectives in Astroparticle and Neutrino THEory with Old and New messengers'' under the program PRIN 2022 funded by the Italian Ministero dell’Università e della Ricerca (MUR).This paper is based upon work supported by the U. S. National Science Foundation (NSF) (grants No. PHY-0919363, No. PHY-1004054, No. PHY-1004072, No. PHY-1242585, No. PHY-1314483, No. PHY-1314507, No. PHY-2310091, associated collaborative grants No. PHY-1211308, No. PHY-1314501, No. PHY-1455351 and No. PHY-1606912, as well as Major Research Instrumentation grant No. MRI-1429544), the Italian Istituto Nazionale di Fisica Nucleare (INFN) (grants from Italian Ministero dell’Università e della Ricerca Progetto Premiale 2013 and Commissione Scientifica Nazionale II), the Natural Sciences and Engineering Research Council of Canada, SNOLAB, and the Arthur B. McDonald Canadian Astroparticle Physics Research Institute. We also acknowledge the financial support by LabEx UnivEarthS (ANR-10-LABX-0023 and ANR18-IDEX-0001), Chinese Academy of Sciences (113111KYSB20210030) and National Natural Science Foundation of China (12020101004).This work has also been supported by the Sao Paulo Research Foundation (FAPESP) grant No. 2021/11489-7. I. Albuquerque and E.M. Santos are partially supported by the National Council for Scientific and Technological Development (CNPq). Some authors were also supported by the Spanish Ministry of Science and Innovation (MICINN) through grant No. PID2019-109374GB-I00, the “Atraccion de Talento” grant No. 2018-T2/TIC-10494, the Polish NCN grant No. UMO-2023/51/B/ST2/02099 and UMO-2022/47/B/ST2/02015, the Polish Ministry of Science and Higher Education (MniSW) grant No. 6811/IA/SP/2018, the International Research Agenda Programme AstroCeNT grant No. MAB/2018/7 funded by the Foundation for Polish Science from the European Regional Development Fund, the European Union’s Horizon 2020 research and innovation program under grant agreement No. 952480 (DarkWave), the Science and Technology Facilities Council, part of the United Kingdom Research and Innovation, and The Royal Society (United Kingdom), and IN2P3-COPIN consortium grant No. 20-152. We also wish to acknowledge the support from Pacific Northwest National Laboratory, which is operated by Battelle for the U.S. Department of Energy under contract No. DE-AC05-76RL01830. This research was supported by the Fermi National Accelerator Laboratory (Fermilab), a U.S. Department of Energy, Office of Science, HEP User Facility. At the time of this work, Fermilab was managed by Fermi Research Alliance, LLC (FRA), acting under contract No. DE-AC02-07CH11359. 


\bibliography{Bibliography}
\end{document}

%% file: authors_DS50.tex
\author{P.~Agnes}\affiliation{\AQLNGS}\affiliation{\AQGSSI}
\author{I.~F.~Albuquerque}\affiliation{\USP}
\author{T.~Alexander}\affiliation{\PNNL}
\author{A.~K.~Alton}\affiliation{\Augustana}
\author{M.~Ave}\affiliation{\USP}
\author{H.~O.~Back}\affiliation{\PNNL}
\author{G.~Batignani}\affiliation{\PIUniPHY}\affiliation{\PIINFN}
\author{K.~Biery}\affiliation{\FNAL}
\author{V.~Bocci}\affiliation{\RMUnoINFN}
\author{W.~M.~Bonivento}\affiliation{\CAINFN}
% ...

%% file: inst.tex
\newcommand{\Alberta}{Department of Physics, University of Alberta, Edmonton, AB T6G 2R3, Canada}
\newcommand{\APC}{APC, Universit\'e de Paris, CNRS, Astroparticule et Cosmologie, Paris F-75013, France}
\newcommand{\AQLNGS}{INFN Laboratori Nazionali del Gran Sasso, Assergi (AQ) 67100, Italy}
\newcommand{\AQGSSI}{Gran Sasso Science Institute, L'Aquila 67100, Italy}
\newcommand{\AQUni}{
Department of Industrial and Information Engineering and Economics, Università degli Studi dell'Aquila, L'Aquila 67100, Italy}
\newcommand{\AstroCeNT}{AstroCeNT, Nicolaus Copernicus Astronomical Center of the Polish Academy of Sciences, 00-614 Warsaw, Poland}
\newcommand{\Augustana}{Physics Department, Augustana University, Sioux Falls, SD 57197, USA}
\newcommand{\Belgorod}{Radiation Physics Laboratory, Belgorod National Research University, Belgorod 308007, Russia}
\newcommand{\BHSU}{School of Natural Sciences, Black Hills State University, Spearfish, SD 57799, USA}
\newcommand{\BINP}{Budker Institute of Nuclear Physics, Novosibirsk 630090, Russia}
\newcommand{\Birmingham}{School of Physics and Astronomy, University of Birmingham, Edgbaston, B15 2TT, Birmingham, UK}
\newcommand{\BNLaddress}{Brookhaven National Laboratory, Upton, NY 11973, USA}
\newcommand{\BOINFN}{INFN Bologna, Bologna 40126, Italy}
\newcommand{\BOUniPHY}{Department of Physics and Astronomy, Universit\`a degli Studi di Bologna, Bologna 40126, Italy}
\newcommand{\CAUniCHE}{Department of Mechanical, Chemical, and Materials Engineering, Universit\`a degli Studi, Cagliari 09042, Italy}
\newcommand{\CAUniEEE}{Department of Electrical and Electronic Engineering, Universit\`a degli Studi di Cagliari, Cagliari 09123, Italy}
\newcommand{\CAUniPHY}{Physics Department, Universit\`a degli Studi di Cagliari, Cagliari 09042, Italy}
\newcommand{\CAINFN}{INFN Cagliari, Cagliari 09042, Italy}
\newcommand{\Carleton}{Department of Physics, Carleton University, Ottawa, ON K1S 5B6, Canada}
\newcommand{\Campinas}{Physics Institute, Universidade Estadual de Campinas, Campinas 13083, Brazil}
\newcommand{\Columbia}{Physics Department, Columbia University, New York, NY 10027, USA}
\newcommand{\Chicago}{Department of Physics and Kavli Institute for Cosmological Physics, University of Chicago, Chicago, IL 60637, USA}
\newcommand{\BOCentroFermi}{Museo Storico della Fisica e Centro Studi e Ricerche Enrico Fermi, Roma 00184, Italy}
\newcommand{\CERNaddress}{CERN, European Organization for Nuclear Research 1211 Geneve 23, Switzerland, CERN}
\newcommand{\CIEMAT}{CIEMAT, Centro de Investigaciones Energ\'eticas, Medioambientales y Tecnol\'ogicas, Madrid 28040, Spain}
\newcommand{\Cluj}{National Institute for R\&D of Isotopic and Molecular Technologies, Cluj-Napoca, 400293, Romania}
\newcommand{\CPPM}{Centre de Physique des Particules de Marseille, Aix Marseille Univ, CNRS/IN2P3, CPPM, Marseille, France}
\newcommand{\CTINFN}{INFN Catania, Catania 95121, Italy}
\newcommand{\CTUNI}{Universit\`a of Catania, Catania 95124, Italy}
\newcommand{\CTLNS}{INFN Laboratori Nazionali del Sud, Catania 95123, Italy}
\newcommand{\ENSMP}{\'Ecole nationale sup\'erieure des mines de Paris, Paris 75272, France}
\newcommand{\ENUniCEE}{Engineering and Architecture Department, Universit\`a di Enna Kore, Enna 94100, Italy}
\newcommand{\ETHZ}{Institute for Particle Physics and Astrophysics, ETH Zurich, Zurich 8093, Switzerland}
\newcommand{\FNAL}{Fermi National Accelerator Laboratory, Batavia, IL 60510, USA}
\newcommand{\FortLewis}{Department of Physics and Engineering, Fort Lewis College, Durango, CO 81301, USA}
\newcommand{\GEUni}{Physics Department, Universit\`a degli Studi di Genova, Genova 16146, Italy}
\newcommand{\GEINFN}{INFN Genova, Genova 16146, Italy}
\newcommand{\Hawaii}{Department of Physics and Astronomy, University of Hawai'i, Honolulu, HI 96822, USA}
\newcommand{\Houston}{Department of Physics, University of Houston, Houston, TX 77204, USA}
\newcommand{\IHEP}{Institute of High Energy Physics, Chinese Academy of Sciences, Beijing 100049, China}
\newcommand{\INFN}{Istituto Nazionale di Fisica Nucleare, Roma 00186, Italia}
\newcommand{\IPNO}{Institut de Physique Nucl\`eaire d'Orsay, 91406, Orsay, France}
\newcommand{\INSTM}{Interuniversity Consortium for Science and Technology of Materials, Firenze 50121, Italy}
\newcommand{\IPHC}{IPHC, Universit\'e de Strasbourg, CNRS/IN2P3, Strasbourg 67037, France}
\newcommand{\JINR}{Joint Institute for Nuclear Research, Dubna 141980, Russia}
\newcommand{\Krakow}{M.~Smoluchowski Institute of Physics, Jagiellonian University, 30-348 Krakow, Poland}
\newcommand{\Kurchatov}{National Research Centre Kurchatov Institute, Moscow 123182, Russia}
\newcommand{\Laurentian}{Department of Physics and Astronomy, Laurentian University, Sudbury, ON P3E 2C6, Canada}
\newcommand{\Lancaster}{Physics Department, Lancaster University, Lancaster LA1 4YB, UK}
\newcommand{\Liverpool}{Department of Physics, University of Liverpool, The Oliver Lodge Laboratory, Liverpool L69 7ZE, UK}
\newcommand{\LNFINFN}{INFN Laboratori Nazionali di Frascati, Frascati 00044, Italy}
\newcommand{\LNLINFN}{INFN Laboratori Nazionali di Legnaro, Legnaro (Padova) 35020, Italy}
\newcommand{\Lodz}{Institute of Applied Radiation Chemistry, Lodz University of Technology, 93-590 Lodz, Poland}
\newcommand{\LPNHE}{LPNHE, CNRS/IN2P3, Sorbonne Universit\'e, Universit\'e Paris Diderot, Paris 75252, France}
\newcommand{\Mainz}{Institut f\"ur Kernphysik, Johannes Gutenberg-Universit\"at Mainz, D-55128 Mainz, Germany}
\newcommand{\Manchester}{Department of Physics and Astronomy, The University of Manchester, Manchester M13 9PL, UK}
\newcommand{\MEPhI}{National Research Nuclear University MEPhI, Moscow 115409, Russia}
\newcommand{\MendeleevUniverisity}{Mendeleev University of Chemical Technology, Moscow 125047, Russia}
\newcommand{\MIBIINFN}{INFN Milano Bicocca, Milano 20126, Italy}
\newcommand{\MIINFN}{INFN Milano, Milano 20133, Italy}
\newcommand{\MIPoliICA}{Civil and Environmental Engineering Department, Politecnico di Milano, Milano 20133, Italy}
\newcommand{\MIPoliCHE}{Chemistry, Materials and Chemical Engineering Department ``G.~Natta", Politecnico di Milano, Milano 20133, Italy}
\newcommand{\MIPoliEIB}{Electronics, Information, and Bioengineering Department, Politecnico di Milano, Milano 20133, Italy}
\newcommand{\MIPoliENE}{Energy Department, Politecnico di Milano, Milano 20133, Italy}
\newcommand{\MIUni}{Physics Department, Universit\`a degli Studi di Milano, Milano 20133, Italy}
\newcommand{\MSU}{Skobeltsyn Institute of Nuclear Physics, Lomonosov Moscow State University, Moscow 119234, Russia}
\newcommand{\NAINFN}{INFN Napoli, Napoli 80126, Italy}
\newcommand{\NAUniPHY}{Physics Department, Universit\`a degli Studi ``Federico II'' di Napoli, Napoli 80126, Italy}
\newcommand{\NAUniCHE}{Chemical, Materials, and Industrial Production Engineering Department, Universit\`a degli Studi ``Federico II'' di Napoli, Napoli 80126, Italy}
\newcommand{\NAUniDIST}{Department of Structures for Engineering and Architecture, Universit\`a degli Studi ``Federico II'' di Napoli, Napoli 80126, Italy}
\newcommand{\NAUniPHARM}{Pharmacy Department, Universit\`a degli Studi ``Federico II'' di Napoli, Napoli 80131, Italy}
\newcommand{\NAUniStruct}{Department of Strutture per l'Ingegneria e l'Architettura, Universit\`a degli Studi ``Federico II'' di Napoli, Napoli 80131, Italy}
\newcommand{\NAUniEEIT}{Department of Electrical Engineering and Information Technology, Universit\`a degli Studi ``Federico II'' di Napoli, Napoli 80125, Italy}
\newcommand{\NSU}{Novosibirsk State University, Novosibirsk 630090, Russia}
\newcommand{\OACINAF}{INAF Osservatorio Astronomico di Capodimonte, 80131 Napoli, Italy}
\newcommand{\Oxford}{University of Oxford, Oxford OX1 2JD, United Kingdom}
\newcommand{\Petersburg}{Saint Petersburg Nuclear Physics Institute, Gatchina 188350, Russia}
\newcommand{\PGUniCBB}{Chemistry, Biology and Biotechnology Department, Universit\`a degli Studi di Perugia, Perugia 06123, Italy}
\newcommand{\PGINFN}{INFN Perugia, Perugia 06123, Italy}
\newcommand{\PIINFN}{INFN Pisa, Pisa 56127, Italy}
\newcommand{\PIUniPHY}{Physics Department, Universit\`a degli Studi di Pisa, Pisa 56127, Italy}
\newcommand{\PNNL}{Pacific Northwest National Laboratory, Richland, WA 99352, USA}
\newcommand{\Princeton}{Physics Department, Princeton University, Princeton, NJ 08544, USA}
\newcommand{\Queens}{Department of Physics, Engineering Physics and Astronomy, Queen's University, Kingston, ON K7L 3N6, Canada}
\newcommand{\RHUL}{Department of Physics, Royal Holloway University of London, Egham TW20 0EX, UK}
\newcommand{\RMTreINFN}{INFN Roma Tre, Roma 00146, Italy}
\newcommand{\RMTreUni}{Department of Mathematics and Physics – Roma Tre University, Roma 00146, Italy}
\newcommand{\RMUnoINFN}{INFN Sezione di Roma, Roma 00185, Italy}
\newcommand{\RMUnoUni}{Physics Department, Sapienza Universit\`a di Roma, Roma 00185, Italy}
\newcommand{\SAINFN}{INFN Salerno, Salerno 84084, Italy}
\newcommand{\SNL}{Savannah River National Laboratory, Jackson, SC 29831, United States}
\newcommand{\SAUni}{Physics Department, Universit\`a degli Studi di Salerno, Salerno 84084, Italy}
\newcommand{\SNOLAB}{SNOLAB, Lively, ON P3Y 1N2, Canada}
\newcommand{\SSUniCHP}{Chemistry and Pharmacy Department, Universit\`a degli Studi di Sassari, Sassari 07100, Italy}
\newcommand{\STFCInterconnect}{Science \& Technology Facilities Council (STFC), Rutherford Appleton Laboratory, Technology, Harwell Oxford, Didcot OX11 0QX, UK}
\newcommand{\STFCppd}{Science \& Technology Facilities Council (STFC), Rutherford Appleton Laboratory, Particle Physics Department, Harwell Oxford, Didcot OX11 0QX, UK}
\newcommand{\Sussex}{Physics and Astronomy Department, University of Sussex, Brighton BN1 9QH, UK}
\newcommand{\Temple}{Physics Department, Temple University, Philadelphia, PA 19122, USA}
\newcommand{\TNFBK}{Fondazione Bruno Kessler, Povo 38123, Italy}
\newcommand{\TNTIFPA}{Trento Institute for Fundamental Physics and Applications, Povo 38123, Italy}
\newcommand{\TNUni}{Physics Department, Universit\`a degli Studi di Trento, Povo 38123, Italy}
\newcommand{\TOINFN}{INFN Torino, Torino 10125, Italy}
\newcommand{\TOPoli}{Department of Electronics and Telecommunications, Politecnico di Torino, Torino 10129, Italy}
\newcommand{\TOUni}{Physics Department, Universit\`a degli Studi di Torino, Torino 10125, Italy}
\newcommand{\TRIUMF}{TRIUMF, 4004 Wesbrook Mall, Vancouver, BC V6T 2A3, Canada}
\newcommand{\TUM}{Physik Department, Technische Universit\"at M\"unchen, Munich 80333, Germany}
\newcommand{\UB}{Universiatat de Barcelona, Barcelona E-08028, Catalonia, Spain} 
\newcommand{\UCDavis}{Department of Physics, University of California, Davis, CA 95616, USA}
\newcommand{\UCRiverside}{Department of Physics and Astronomy, University of California, Riverside, CA 92507, USA}
\newcommand{\UCSanDiego}{Department of Physics, University of California, San Diego, CA 92093, USA}
\newcommand{\UCLA}{Physics and Astronomy Department, University of California, Los Angeles, CA 90095, USA}
\newcommand{\UCAS}{University of Chinese Academy of Sciences, Beijing 100049, China}
\newcommand{\UMass}{Amherst Center for Fundamental Interactions and Physics Department, University of Massachusetts, Amherst, MA 01003, USA}
\newcommand{\UNAM}{Instituto de F\'isica, Universidad Nacional Aut\'onoma de M\'exico, M\'exico 01000, Mexico}
\newcommand{\UnivAQ}{Universit\`a degli Studi dell’Aquila, L’Aquila 67100, Italy}
\newcommand{\UniversityofEdinburgh}{School of Physics and Astronomy, University of Edinburgh, Edinburgh EH9 3FD, UK}
\newcommand{\UOC}{Department of Chemistry, University of Crete, P.O. Box 2208, 71003 Heraklion, Crete, Greece}
\newcommand{\USP}{Instituto de F\'isica, Universidade de S\~ao Paulo, S\~ao Paulo 05508-090, Brazil}
\newcommand{\VTech}{Virginia Tech, Blacksburg, VA 24061, USA}
\newcommand{\Washington}{Center for Experimental Nuclear Physics and Astrophysics, and Department of Physics, University of Washington, Seattle, WA 98195, USA}
\newcommand{\Warwick}{University of Warwick, Department of Physics, Coventry CV47AL, UK}
\newcommand{\WUT}{Institute of Radioelectronics and Multimedia Technology, Faculty of Electronics and Information Technology, Warsaw University of Technology, 00-661 Warsaw, Poland}
\newcommand{\WilliamsCollege}{Williams College, Department of Physics and Astronomy, Williamstown, MA 01267 USA}
\newcommand{\Zaragoza}{Centro de Astropart\'iculas y F\'isica de Altas Energ\'ias, Universidad de Zaragoza, Zaragoza 50009, Spain}
\newcommand{\ZaragozaARAID}{Fundaci\'on ARAID, Universidad de Zaragoza, Zaragoza 50009, Spain}
\newcommand{\UniHAM}{Institute of Experimental Physics, University of Hamburg, Luruper Chaussee 149, 22761, Hamburg, Germany}
\newcommand{\SDakota}{School of Natural Sciences, Black Hills State University, Spearfish, South Dakota 57799, USA}
\newcommand{\Saclay}{Universit\'{e} Paris-Saclay, CEA, List, Laboratoire National Henri Becquerel (LNE-LNHB),
F-91120 Palaiseau, France}
\newcommand{\NASS}{Scuola Superiore Meridionale,  80138 Napoli, Italy}